\documentclass{aa}  

\usepackage{graphicx}
\usepackage{txfonts}
\newcommand{\lppr}{\stackrel{<}{\scriptstyle \sim}}
\begin{document}

   \title{Rotationally-driven VHE emission from the Vela pulsar}

   \author{Z. Osmanov
          \inst{1}
          \and
          F. M. Rieger\inst{2,3}
          }

   \institute{School of Physics, Free University of Tbilisi, 0183, Tbilisi, Georgia;
              \email{z.osmanov@astro-ge.org}
        \and
             Zentrum f\"ur Astronomie (ZAH), Institut f\"ur Theoretische Astrophysik, Universit\"at
             Heidelberg, Philosophenweg 12, 69120 Heidelberg; 
             \email{f.rieger@uni-heidelberg.de}
        \and     
              Max-Planck-Institut f\"ur Kernphysik, P.O. Box 103980, 69029 Heidelberg, Germany
             }

   \date{Received ; accepted }

 
  \abstract
   {The recent detection of pulsed $\gamma$-ray emission from the Vela pulsar in 
   the $\sim10$ to 100 GeV range by H.E.S.S. promises an important potential to 
   probe into the very high energy (VHE) radiation mechanisms of pulsars.}
   {A combined analysis of H.E.S.S. and Fermi-LAT data suggests that the leading 
   wing of the P2 peak shows a new, hard gamma-ray component (with photon index 
   as hard as $\Gamma \sim 3.5$), setting in above 50 GeV and extending beyond 
   100 GeV. We study these findings in the context of rotationally-driven 
   (centrifugal) particle acceleration.}
   {We analyze achievable particle energies in the magnetosphere of the Vela 
   pulsar, and calculate the resultant emission properties.}
   {Inverse Compton up-scattering of thermal photons from the surface of the star 
   is shown to lead a pulsed VHE contribution reaching into the TeV regime with 
   spectral characteristics compatible with current findings. If confirmed by 
   further observations this could be the second case where rotationally-driven 
   processes turn out to be important to understand the VHE emission in young 
   pulsars.}
   {}


   \maketitle
%

\section{Introduction}

One of the most enigmatic high energy astrophysical objects is the Vela pulsar (PSR B0833-45). 
Located at a distance of $d_l\simeq 300$ pc \citep{Caraveo2001} the Vela pulsar is a young pulsar 
with a period of rotation of $P=89$ ms, an estimated age of $t =P/(2\dot{P}) \simeq 1.1\times 
10^4$ yr \citep{Reichley1970} and a spin-down luminosity of the order of $\dot{E}\simeq 10^{37}$ 
erg/sec \citep{Manchester2005,Mignani2017}.

Pulsed gamma-ray emission from the MeV to GeV domain has been reported by a variety of instruments 
early on, from SAS-2 and COS B to CGRO, AGILE, {\it Fermi}-LAT \citep{vela1,vela2,vela3,egret,agile,fermilat1}.
At high $\gamma$-ray energies its light curve is found to exhibit two prominent peaks, P1 and P2, which 
are separated in phase by $0.43$ and connected by a so-called P3 bridge.

In particular, while {\it Fermi}-LAT initially detected high energy (HE; $>100$ MeV) emission from 
the Vela pulsar up to $\sim10$ GeV, further observations have increased this to $20$ GeV 
\citep{fermilat2} and $50$ GeV \citep{leung}, respectively. At very high energies (VHE; $>100$ GeV), 
on the other hand, the Vela pulsar has not been seen in early H.E.S.S. observations with an energy 
threshold of $170$ GeV, imposing constraints on a possible inverse Compton (IC) contribution from 
outer gap models \citep{hess1}. New results based on H.E.S.S. II (CT5-only) observations in the 
sub-20 GeV to 100 GeV range have been recently reported \citep{hess2}, with pulsed $\gamma$-rays 
from the strong P2 peak (phase interval $0.5-0.6$) detected at a significance level of $15\sigma$. 
The results for P2 confirm the sub-exponential cutoff shape found at lower energies with 
{\it Fermi}-LAT and support the presence of emission above 100 GeV. Perhaps most intriguingly, there 
are converging indications for the emergence of a hard spectral component above 50 GeV in the leading 
wing of the second peak, LW2 (phase interval $0.45-0.5$), which possibly extends beyond 100 GeV. 
A simple power law fit of the H.E.S.S. II data for LW2 with two different energy thresholds, CI and 
CII, yields comparable photon indices of $\Gamma = 3.72\pm 0.51$ (CI) and $\Gamma = 3.48\pm 0.21$ 
(CII), respectively \citep{hess2}. Here CI and CII refer to different analysis cuts optimized such as to 
yield a large effective area in the 10-20 GeV range, with the threshold for CI approaching 5-10 GeV 
depending on zenith angles, and the threshold being approximately two times higher for CII.
If the inferred hardening is confirmed by further observations it would point to the presence of a second 
and new emission component.

Motivated by these indications we study the possible generation of high and very high energy 
gamma-rays in the context of magneto-centrifugal acceleration \citep[e.g.,][]{macrog,Gangadhara1996,
or09,or17,Bogovalov2014}. 
As the magnetic field in the magnetospheres of pulsars is extremely high, charged particles 
will quickly transit to the ground Landau level and start sliding along the magnetic field 
lines. Since the field is co-rotating with the star, the particle dynamics is likely to be 
highly influenced by the effects of centrifugal acceleration. This particularly applies for 
particles approaching the light cylinder (LC) surface. 

The possible relevance of this for the origin of the emission in the Crab pulsar has been 
studied early on by \cite{gold}, suggesting that this could facilitate an efficient energy 
transfer from the rotator into the kinetic energy of plasma particles. Using an idealized
model for centrifugal acceleration, \cite{macrog} later on showed that incorporation of
(special) relativistic effects can lead to a radial deceleration of the particle motion,  
resembling features known from general relativity \citep{Abramowicz1990}.

As a convenient analogy, centrifugal particle acceleration has since then been studied in 
a variety of contexts, e.g. for predicting the location to frequency mapping in pulsars 
\citep{gang1,gang2} or for investigating the origin of the gamma-ray emission from the 
rotating jet base in active galactic nuclei (AGN) \citep[e.g.,][]{ganglesch,riegman,
Xu02,orb07,ghis09,Osmanov2014}. In particular, applying magneto-centrifugal acceleration 
to Crab-like pulsars it was found that on approaching the LC area electrons could achieve 
Lorentz factor up to $\gamma\sim 10^7$ \citep[][]{or09,Bogovalov2014}, suggesting that 
inverse Compton (IC) scattering in the Crab Pulsar's magnetosphere could generate detectable 
pulsed VHE emission in the TeV band. A subsequent evaluation of the spectral characteristics 
\citep{or17} turned out to be in a good agreement with the observational data.

The present paper focuses on an application to the Vela pulsar, assessing the potential of magneto-centrifugal 
acceleration to generate detectable VHE gamma-ray emission. As we show in 
Sec.~2 inverse Compton (IC) up-scattering of thermal soft photons can indeed facilitate the production 
of pulsed VHE emission at levels compatible with current experimental findings. This allows for 
interesting insights into pulsar physics.

\section[]{Rotational-driven $\gamma$-ray emission}
For simplicity we consider a single particle approach and assume that the magnetospheric electrons 
follow the co-rotating magnetic field lines. In particular, the Vela pulsar is characterized by 
$P= 0.089$ sec and $dP/dt\approx 1.25\times 10^{-13}$ ss$^{-1}$. This leads to a strong magnetic 
induction close to the star's surface of $B_{\rm st}\approx 3.2\times 10^{19} \sqrt{P\dot{P}} \approx 
3.4\times 10^{12}$ Gauss. Therefore, plasma particles will follow the field lines, accelerating 
along them. For a single relativistic massive particle one can define the Lagrangian 
\begin{equation}
\label{lag} L = -m\left(1-\frac{\upsilon^2}{c^2}-\frac{\omega^2r^2\sin^2\theta}{c^2}\right)^{1/2},
\end{equation}
where $m$ is the rest mass of the particle, $\upsilon=\dot{r}$, $r$ is the radial coordinate along 
the straight magnetic field line inclined by $\theta$ with respect to the rotation axis and $\omega =
 2\pi/P$ is the angular velocity of rotation. Since $L$ does not depend explicitly on time, the 
corresponding Hamiltonian, $H = \dot{r}\frac{\partial L}{\partial r}-L$, is a constant of motion, 
resulting in the following expression for the particle Lorentz factor \citep{or17}
\begin{equation}
\label{gamma} \gamma(r) = \gamma_0\,\frac{1-r_0^2/r_{l}^2}{1-r^2/r_{l}^2},
\end{equation}
where by $r_0$ and $\gamma_0$ we denote respectively the initial radial coordinate and Lorentz factor, 
$r_l=R_{lc}/\sin\theta$ and $R_{lc}=c/\omega=4.25\times 10^8$ cm is the LC radius. As evident from 
Eq.~(\ref{gamma}), on approaching the LC area the Lorentz factor asymptotically increases, leading to 
a very efficient acceleration process. Close to the LC zone the corresponding acceleration time-scale, 
$t_{\rm acc}=\gamma/\dot{\gamma}$, behaves as 
\citep{or17}
\begin{equation}
\label{acc} t_{\rm acc}(\gamma)\approx \frac{R_{lc}}{2c\sin\theta}\left(1-\frac{r_0^2\sin^2\theta}{R_{lc}^2}\right)^{1/2}
\left(\frac{\gamma_0}{\gamma}\right)^{1/2}.
\end{equation}
This time-scale is a continuously decreasing function of $\gamma$, and since the latter formally tends 
to infinity on the LC, $t_{acc}$ tends to zero, indicating an extremely efficient process.

\begin{figure}
  \resizebox{\hsize}{!}{\includegraphics[angle=0]{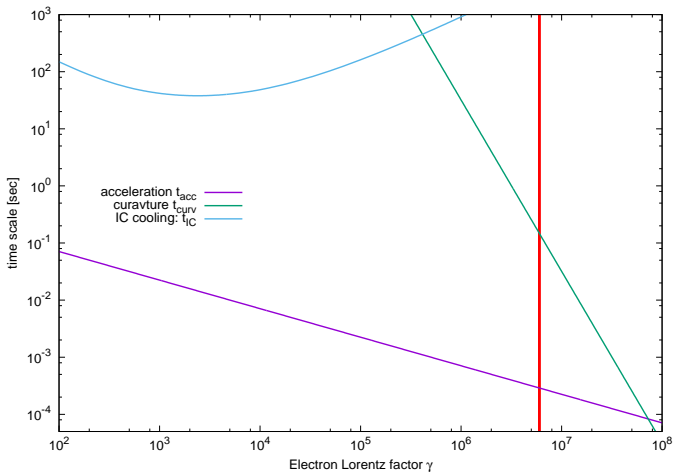}}
  \caption{Characteristic time-scales for the Vela pulsar as a function of particle Lorentz factor: centrifugal acceleration time-scale 
  (pink), curvature cooling time-scale (green) and IC cooling time-scale (blue). The red vertical line represents the upper limit 
  imposed by $\gamma_{\rm cor}$. Acceleration here is essentially limited by $\gamma_{\rm cor}$. The set of parameters employed
  is $\gamma_0 = 10^4$, $\gamma_{\rm cor}= 6\times 10^6$ and $\sin\theta\sim 1$.} \label{fig1}
\end{figure}

It is clear that acceleration cannot be a continuous process and must be terminated. As already mentioned, 
centrifugal acceleration is provided by the frozen-in condition of magnetospheric plasma particles. This condition 
will only last until the plasma energy density, $n_{\rm GJ}(r)\gamma mc^2$, becomes comparable to the energy 
density in the magnetic field $B^2/8\pi$, where $n_{\rm GJ}(r)={\bf \omega \cdot B}\,M\left(2\pi ec\right)^{-1}=
M\, n_{0,{\rm GJ}}$ is the Goldreich-Julian particle number density \citep{GJ}, taking account of the multiplicity factor, 
$M=\left(1-r^2/r_l^2\right)^{-1}$. Assuming a dipolar magnetic field structure, $B(r_l)\simeq B_{\rm st}\times (r_s/r_l)^3$, 
with $r_s=11.5$ km \citep[e.g.,][]{Li2016}, the co-rotation condition imposes a constraint on the maximum attainable 
electron Lorentz factor of
\begin{equation}\label{gcor} 
\gamma_{\rm cor}\approx 6.4\times 10^6\left(\frac{\gamma_0}{10^4}\right)^{1/2}\left(\frac{\sin^3\theta}{\cos\theta}\right)^{1/2},
\end{equation}
taking Lorentz factors $\gamma_0\sim 10^4$ for the secondaries \citep{daugherty} as reference value. 
Similar to the emission model proposed by \cite{gold} electrons close to the LC area co-rotate with 
almost $R_{lc}$, suggesting that curvature $\gamma$-ray emission might be of significance. The 
corresponding curvature photon energy $3hc\gamma^3/(4\pi R_{lc})$ is in the range
\begin{equation}\label{ecur} 
\epsilon_{\rm cur}\approx 8.7\times\left(\frac{5\times 10^6}{\gamma}\right)^3 \mathrm{MeV}\,.
\end{equation}
Hence curvature emission could in principle lead to a pulsed $\gamma$-ray contribution in the COMPTEL 
(1-10) MeV regime. It is clear that curvature emission also introduces some constraints on the maximum 
attainable energies. The cooling time-scale due to curvature emission, $t_{\rm cur}=\gamma mc^2/P_{\rm 
cur}$ where $P_{\rm cur}= 2e^2c\gamma^4/(3R_{lc}^2)$ is the single particle power \citep{Ochelkov1980}, 
becomes
\begin{equation}\label{cur} 
t_{\rm cur}\approx 0.25\times\left(\frac{5\times10^6}{\gamma}\right)^3 \mathrm{sec}\,.
\end{equation}
Acceleration would thus formally be balanced by curvature cooling at 
\begin{equation}\label{gcur} 
\gamma_{\rm cur}\approx 7.2\times 10^7\left(\frac{10^4}{\gamma_0}\right)^{1/5}\sin^{2/5}\theta\,,
\end{equation}
assuming $r_0<<R_{lc}$. Hence the co-rotation limit (see Eq.~(\ref{gcor})) is generally expected to 
impose the most stringent constraints, see also Fig.~\ref{fig1} for illustration.

The above analysis suggests that rotational-driven curvature emission could lead to a non-negligible 
$\gamma$-ray contribution at $\sim (1-10)$ MeV energies, that could help alleviating tensions in e.g. 
outer-gap SED modelling of the Vela pulsar \cite[e.g.,][]{Takata2004}. One can roughly estimate 
the possible curvature MeV contribution from $L_{cur}\simeq 2\, n_{\rm 0,GJ} M  \Delta V P_{\rm cur}$ 
(see also eqs.~[\ref{d}] and [\ref{IC_lum}]), yielding $L_{cur} \sim 10^{31}\,(B/6\times 10^4 
\mathrm{G})\,(\gamma/4\times 10^6)^4\, \xi^2 \chi \,\cos\alpha$ erg/s.
In the COMPTEL regime the flux  is of the order of $F_{_{COM}}\sim 10^{-10}$erg s$^{-1}$ cm$^{-2}$ 
\citep{comptel}, translating into an isotropic-equivalent luminosity $L_i\simeq 4\pi d_l^2 F_{_{COM}}\simeq 
10^{33}$ erg s$^{-1}$. If one assumes that focusing in this energy band is of the same order as in TeV 
regime, $0.25$sr (see the estimates below), the inferred luminosity $L\sim L_i\left(0.25/4\pi\right)\sim 
10^{31}$ erg s$^{-1}$ would be very close to $L_{cur}$ derived above.

On the other hand, as shown below, rotational-driven IC scattering of thermal soft photons from the surface 
of the neutron star can lead to a VHE contribution in the $\sim(0.1-1)$ TeV regime. This is different to typical outer 
gap models, that would predict a rising IC contribution peaking at $\sim 10$ TeV \citep[e.g.,][]{Aharonian2003,Takata2006}.\\
The surface temperature of the Vela pulsar is below the standard cooling curve and probably in the 
range $T\simeq (6-8)\times 10^5$ K \citep{Page1996,Mori2004,weisskopf}. We take $T=7\times 10^5$ K 
as reference value in the following. The thermal photons thus have characteristic peak energies of 
the order of $\epsilon_{ph}\approx 2.8 kT\sim 0.2$ keV, where $k$
is the Boltzmann constant. If these photons encounter centrifugally-accelerated electrons the former 
will inevitably gain energy. Since the ratio $\gamma\epsilon_{ph}/(mc^2)$ exceeds unity for electron 
Lorentz factors exceeding $\gamma \approx 2.3\times 10^3$, IC up-scattering to the VHE regime essentially occurs 
in the Klein-Nishina (KN) regime where the corresponding single particle IC power can be 
suitably approximated by \citep[e.g.,][]{blum}
\begin{equation}\label{PKN}
    P_{\rm IC,KN}\simeq\frac{\sigma_T\left(mckT\right)^2}{16\hbar^3}\left(\ln
                \frac{4\gamma kT}{mc^2}-1.981\right)\left(\frac{r_s}{r_l}\right)^2\,,
\end{equation} 
with $\sigma_T$ the Thomson cross section and $\hbar$ the Planck constant.  We use the general expression 
for the electron loss rate derived in \cite{Aharonian1981} to evaluate the corresponding IC cooling time-scale 
$t_{\rm IC} = \gamma mc^2/P_{\rm IC}$. The result is shown in Fig.~\ref{fig1} together with the time-scales for 
curvature losses $t_{\rm curv}$ and particle acceleration $t_{\rm acc}$, respectively.
As evident from this figure IC scattering does not impose any constraints on the maximum attainable 
energies of electrons, the latter essentially being determined by the co-rotation limit.

Despite the negligible role of IC in limiting maximum attainable energies, up-scattering of thermal 
photons could lead to detectable $\gamma$-ray emission in the VHE energy regime \cite[cf. also][]{Bogovalov1992}. 
Scattered (KN) photon energies,  
\begin{equation}\label{EKN}
    \epsilon_{\gamma} \sim\gamma mc^2\sim 100\times \left(\frac{\gamma}{2\times 10^5}\right)\, 
    \mathrm{GeV},
\end{equation} may in principle extend into the TeV regime (for $\gamma\leq\gamma_{\rm cor}$), 
though expected TeV flux levels could be too low to allow for a significant detection. The latest 
H.E.S.S. observations of the Vela pulsar in fact provide indications for $\gamma$-ray emission above 
100 GeV \citep{hess2}. For an order of magnitude estimate of the intrinsic rotational-driven IC-VHE 
(100 GeV) luminosity we take into account that the process of efficient acceleration and radiation 
production occurs in a thin shell close to the LC \citep{or09} whose characteristic width is given 
by
\begin{equation}\label{d} 
d\approx \left|\frac{\gamma}{d\gamma/dr}\right| \approx R_{lc}\frac{\gamma_0}{2\gamma}\,.
\end{equation}
Multiplying the corresponding particle number with the single IC power $P_{\rm IC, KN}$, Eq.~(\ref{PKN}), 
one obtains 
\begin{eqnarray}\label{IC_lum}
L_{\rm IC}^{\rm VHE} &\approx& 2\, n_{\rm 0,GJ} M  \Delta V P_{\rm IC, KN}\nonumber\\
                     &\sim&2 \times 10^{29}  \left(\frac{B}{6 \times 10^4~\mathrm{G}}\right)
                                      \xi^2 \chi \,\cos\alpha\,\sin^2\alpha\; \mathrm{erg/s}\,,
                                      \nonumber\\ 
\end{eqnarray} using $M \sim \gamma_{\rm cor}/\gamma_0$ and $\gamma_{\rm cor}/\gamma=30$, 
and denoting by $\Delta V\approx \chi \left(\delta l\right)^2 d$ the corresponding volume; $\chi\lppr1$ 
is a dimensionless factor depending on the topology of magnetic field lines, $\delta l \sim R_{lc} 
\theta$ is the azimuthal length scale involved in this process, and $\theta \sim \Omega P\,(\xi/10)=
2\pi~(\xi /10)$ is the corresponding angle, where $(\xi/10) P$ is an approximate value for the pulse 
duration ($\xi\leq 10$). Obviously, $L_{\rm IC}^{\rm VHE} \ll$ the spin-down luminosity $\dot{E}$.

A rough estimate for the corresponding power in electrons, $\dot{E}_e$, can be obtained from
$\dot{E}_e=\dot{N}_{\rm IC} \gamma_{\rm IC}m_e c^2$. Here, $\dot{N}_{\rm IC}$ is the total rate 
of IC emitting electrons, $\dot{N}_{\rm IC} \propto \dot{N}_0/\gamma_{\rm IC}^{1/2}$ (cf. eq.~[\ref{neq}]), 
and $\dot{N}_0 = n_{\rm 0,GJ}~M~ dA~\dot{r}$ where $dA\simeq  \chi\, R_{lc}^2 \,(2\pi/10)^2\, \xi^2 $ 
using the notation above. Assuming $M\simeq \gamma_{\rm cor}/\gamma_0$ one finds $\dot{E}_e \sim 
10^{32}$ erg/s, again much below the spin-down luminosity $\dot{E}\simeq 10^{37}$ erg/sec of the Vela 
pulsar. In principle, polar cap heating by secondary particles flowing back to the neutron star could
lead to some excess thermal emission and thereby modify the IC emission. As shown by \citet{harding}, 
however,  this effects is expected to be negligible for Vela-type pulsars.

The recent H.E.S.S. analysis suggests isotropic-equivalent VHE (100 GeV) flux levels of the order 
of $L_i\sim10^{31}$ erg/s. This would indicate that the VHE emission of the Vela pulsar is collimated 
with corresponding emission cones confined to solid angle areas of $\Delta\Omega \simeq 4\pi \,
(L_{\rm IC}^{\rm VHE}/L_i) \lppr 0.25$ sr.

The indications for the emergence of a hard spectral component above 50 GeV in LW2 are particularly 
interesting, alluding to the possibility that VHE emission well beyond 100 GeV may be present in the 
Vela pulsar. To investigate the spectral characteristics of rotationally-driven VHE in more detail, 
we follow the description developed in \cite{riegaha08}. Accordingly, assuming electrons to be injected 
at a constant rate, $Q$, into the region of centrifugal acceleration, the differential particle density distribution 
along the magnetic field lines can be written as
\begin{equation}\label{neq}
n_e(\gamma) \simeq \frac{Q\,t_{acc}}{\gamma}~ H(\gamma-\gamma_0)
                  \propto \gamma^{-3/2},
\end{equation}
where the behaviour $t_{acc}\simeq\gamma^{-1/2}$ (see Eq. (\ref{acc})) has been used. The (effective) 
differential number of electrons, $N_e(\gamma) = \frac{dN_e}{d\gamma}$, then becomes \citep{or17}
\begin{equation}\label{distr}
N_e(\gamma) = n_e(\gamma)\Delta V\propto n_e(\gamma)d\propto\gamma^{-5/2},
\end{equation}
taking into account that higher energy particles are accumulated in a narrower region close to the LC, 
with width scaling as $d\propto 1/\gamma$, Eq.~(\ref{d}). IC up-scattering of thermal soft photons 
in the KN regime then produces a photon spectrum \citep{blum}
\begin{equation}
\frac{dN_{\gamma}}{d\epsilon_{\gamma} dt} \propto
\epsilon_{\gamma}^{-3.5} \left(\ln\frac{\epsilon_{\gamma}
kT}{m^2c^4}-1.4\right) \sim \epsilon_\gamma^{-3.5}
\end{equation} with VHE photon index compatible with the ones inferred for LW2, i.e. $\Gamma= 3.72
\pm 0.51$ (CI) and $\Gamma = 3.48\pm 0.21$ (CII).

\begin{figure}
  \resizebox{\hsize}{!}{\includegraphics[angle=0]{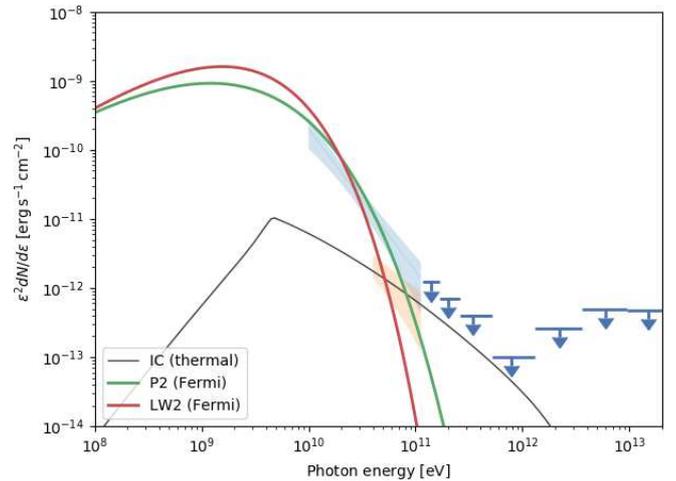}}
  \caption{SED representation with best fits to the phase-resolved spectra based on $\sim8$ year of
  {\it Fermi}-LAT data (green and red lines) \citep{hess2}, along with a characteristic model 
  calculation for the rotational-driven thermal IC contribution at VHE energies (black curve).
  Blue-shaded area represents results from power-law fits of recent H.E.S.S. II data in the range
  10-110 GeV for P2 \citep{hess2}. Orange-shaded area indicates range of H.E.S.S. II power-law 
  indices inferred for LW2 (CII cuts).
  Blue upper limits denote early H.E.S.S. I constraints above 170 GeV \citep{hess1}. 
  In this case rotationally-driven IC $\gamma$-ray emission below $100$ GeV may become detectable 
  for LW2 but not P2.} \label{fig2}
\end{figure}

Figure~\ref{fig2} shows the $\gamma$-ray spectral energy distribution (SED) for the Vela pulsar along 
with a characteristic model calculation of the expected IC $\gamma$-ray contribution using NAIMA \citep{Zabalza2015}. 
The calculations assume (quasi-isotropic) up-scattering of thermal photons ($T=7\times 10^5$ K) close 
to LC by an exponential cut-off power-law electron distribution, $n(\gamma) \propto \gamma^{-2.5} 
\exp(-[\gamma/\gamma_c]^2)$, with $\gamma_c=6\times 10^6$ and $\gamma_0\simeq10^4$. The Fermi-LAT 
data for P2 and LW2 are illustrated by their best-fit (exponential cutoff power law model) representations 
(green and red curve, respectively), as detailed in \cite{hess2}. 

Similar as in other LC models \cite[e.g.,][]{Cheng1986,Bogovalov1992} we assume that particles leaving 
the acceleration zone no longer make a significant contribution to the pulsed VHE emission. This could 
be related to electrons quickly cooling out of the VHE window due to increased synchro-curvature losses 
in bent fields, along with a reduced focusing of the resultant emission.
In particular, particle escaping from the acceleration zone and experiencing non-zero pitch angles \citep[e.g., 
due to excitation of instabilities close to LC,][]{Machabeli2010} would immediately cool down on characteristic 
length scales $l \sim c t_{\rm cool} \propto \frac{1}{\gamma~\sin^2\theta}<< R_{lc} \gamma_0/\gamma_{\rm cor}$ 
for a wide range of angles. In the present model one can roughly estimate the possible synchrotron contribution 
below a few GeV via $L_{syn}(\gamma)\sim N P_{syn} \sim 2~N~[\gamma^2 \sigma_T  c\,(B^2/8\pi) \sin^2\theta]$, 
where $N\sim \dot{N}_{\rm IC}~t_{\rm cool}$, yielding $L_{syn} \sim 10^{32}$ erg/sec. On the other hand, 
for a more detailed spectral modelling, quasi-linear diffusion and the maintenance of pitch angles would have 
to be incorporated \citep[]{Machabeli2010}, which however is beyond the scope of the current paper.

As can be seen in Figure~\ref{fig2} , rotationally-driven IC $\gamma$-ray emission may reach 
into the TeV regime, but become apparent below $\sim 100$ GeV only for LW2. These results 
motivate detailed VHE studies with higher sensitivity, as will become possible with CTA South 
\citep{Burtovoi2017}, in order to better characterize the $\gamma$-ray emission above $\sim100$ 
GeV. If confirmed by further observations, this would highlight the relevance of incorporating 
rotational-driven effects for our understanding of the $\gamma$-ray emission in young pulsars. 
Taking account of the angular dependency in Eq.~(\ref{gcor}) suggests that rotationally-driven IC 
emission could well yield VHE emission up to several TeV. When compared to the Crab pulsar, where r
otationally-driven curvature emission can lead to a detectable GeV contribution \citep{or17}, 
supporting a smoother GeV-VHE connection, as indeed observed \citep{magic}, rotationally-driven 
curvature emission in the Vela pulsar peaks at MeV energies only (eq.~[\ref{ecur}]); rotationally-driven 
IC VHE component in the Vela pulsar thus appears as a separate, second and new component.

We note that rotational acceleration in principle predicts a decrease in individual pulse width 
with increasing energy. Neglecting local changes in the field topology, the observed pulse duration 
is roughly proportional to $\gamma_{\phi}^{-3}$ \citep{RL}, where $\gamma_{\phi}=
\left(1-\upsilon_{\phi}^2/c^2\right)^{-1/2}$ is the azimuthal Lorentz factor. Taking into account Eq. (\ref{gamma}) 
we approximately have $\gamma_{\phi} \simeq(\gamma/\gamma_0)^{1/2}$. Hence, for VHE emission with 
$\gamma\sim10^6$ and $\gamma_0\sim10^4$ for example, the characteristic observed pulse width could 
be as small as $\sim10^{-3}$. This seems interesting when compared with the inferred Gaussian width of 
$\sim 0.002$ for the additional $\gamma$-ray (phasogram) component in the range (10-80) GeV around 
P2 \citep{hess2}.

\section{Conclusion}
The generation of VHE $\gamma$-rays in fast pulsars is still an open issue \citep[e.g.,][]{Bednarek2012,
Mochol2015,Hirotani2015,or17,Harding2018}. The latest H.E.S.S. observations of the Vela pulsar 
in the sub-$20$ GeV to $100$ GeV regime have now revealed pulsed $\gamma$-rays from its P2 peak at 
high significance, and together with an updated analysis of Fermi-LAT data suggest the emergence of a second 
and new, hard VHE component above 50 GeV in the leading wing (LW2) of P2 \citep{hess2}.

Motivated by this, we have analysed the potential role of centrifugal acceleration in generation of 
pulsed VHE emission in the magnetosphere of the Vela pulsar. Our results show that inverse Compton 
up-scattering of thermal photons by rotationally accelerated electrons close to the light cylinder 
could lead to a hard (yet falling) power-law like VHE contribution that reaches into the TeV regime and 
in the case of LW2 could become apparent above $\sim50$ GeV. Future observations with increased 
sensitivity will be important to better probe its spectral characteristics and to eventually allow a 
clear identification. Along with the Crab pulsar  \citep{or17}, this could be the second case where 
rotational-driven $\gamma$-ray emission could turn out to be important to fully understand the VHE 
emission in young pulsars.

\section*{Acknowledgments}
Useful comments by the referee are gratefully acknowledged. We also thank S. Bogovalov and 
D. Khangulyan for discussions. The research of ZO was supported by a German DAAD 
scholarship within the program Research Stays for University Academics and Scientists, 2018 and was partially supported by Shota Rustaveli National Science Foundation Grant NFR17-587. ZO 
also acknowledges hospitality of the MPIK and the group of F. Aharonian during his visit in 2018. 
FRM acknowledges financial support by a DFG Heisenberg Fellowship under RI 1187/6-1.

\end{document}